\documentclass{aa}  

\usepackage{graphicx, subcaption}
\usepackage{booktabs}
\usepackage{natbib}
\bibpunct{(}{)}{;}{a}{}{,} 
\usepackage{txfonts}
\usepackage{hyperref}
\begin{document} 

\newcommand{\ssin}{_\text{in}} 
\newcommand{\ssout}{_\text{out}} 
\newcommand{\tres}{\texttt{TRES}}

   \title{Massive stellar triples on the edge}

   \subtitle{A numerical study of the evolution and final outcomes of destabilized massive triples.}

   \author{C. W. Bruenech
          \inst{1},
          T. Boekholt\inst{2, 3},
          F. Kummer\inst{1},
          S. Toonen\inst{1}
          }

   \institute{Anton Pannekoek Institute for Astronomy, University of Amsterdam, 1090 GE Amsterdam, The Netherlands,\\
              \email{c.w.bruenech@uva.nl, f.a.kummer@uva.nl, s.g.m.toonen@uva.nl}
         \and
              Rudolf Peierls Centre for Theoretical Physics, Clarendon Laboratory, Parks Road, Oxford, OX1 3PU, UK
         \and
              NASA Ames Research Center, Moffett Field, 94035, CA, USA,\\
              \email{tjarda.boekholt@nasa.gov}
             }

   \date{Received September 15, 1996; accepted March 16, 1997}

 
  \abstract
   {Massive stars reside predominantly in triples or higher-order multiples. Their lives can be significantly affected by three-body interactions, making it an important area of study in the context of massive star evolution.}
   {We intend to provide a statistical overview of the lives and final outcomes of massive triples that are born dynamically stable but become unstable due to evolutionary processes.}
   {A population of initially stable triples with a massive primary star are evolved from the zero-age main sequence using the code \tres, which combines stellar evolution with orbit-averaged dynamics. The triples that become unstable are transferred to a direct N-body code where they are simulated until the system disintegrates. This excludes systems undergoing mass transfer, such that the instability is caused by stellar winds or supernovae explosions. Two suites of N-body simulations are performed; one with gravity as the only interaction, and one with stellar evolution included.}
   {We find that our triples remain on the edge of stability for a long time before disintegrating, making stellar evolution a consequential process during this phase. Eventually the destabilization results in either the ejection of a stellar body or the collision between two components. We find that collisions occur in $35 - 40$\% of systems, with the variation in percentages coming from whether or not stellar evolution is included. The collisions predominantly involve two main sequence stars ($70 - 78$\%) or a main sequence and post-main sequence star ($13 - 28$\%). We estimate a Galactic rate of collisions due to massive triple destabilization at $1.1 - 1.3$ events per Myr. Furthermore, we find that the process of destabilization often ends in the ejection of one of the stellar bodies, specifically for $31 - 40$\% of systems. The ejected bodies have typical velocities of $\sim6$ km/s, with a tail stretching to $10^2$ km/s. If we make the assumption that $20$\% of massive stars are runaway stars, then $0.1$\% of runaways originate from triple destabilization. Overall, our simulations show that triple instability affects approximately $2$\% of massive triples. However, we estimate that up to ten times as many systems can become unstable due to mass transfer in the inner binary, and these systems may end up ejecting bodies at higher velocities.}
   {}


   \maketitle
%

\section{Introduction}

Star systems containing two or more stars are common in the universe. Observations show that about half of solar-type stars reside in binaries or higher order multiples, with both the multiplicity fraction and number of companions increasing with the mass of the star \citep{tokovinin_binaries_2014, moe_mind_2017, offner_origin_2023}. As a consequence, the majority of massive stars ($M \geq 8$ M$_\odot$) are expected to have at least two companions. A significant fraction of these systems are likely to interact during the lifetimes of the stars, which can drastically affect the evolution of the stellar components \citep{sana_binary_2012}.

Stellar objects in binary systems follow fixed orbits as long as the system does not experience any external perturbations or internal changes. In comparison, the addition of a third body introduces effects that can alter the configuration of the system over secular timescales due to purely gravitational interactions \citep{von_zeipel_sur_1910, kozai_secular_1962, lidov_evolution_1962, szebehely_stability_1977, naoz_resonant_2013}. Stellar evolution combined with these three-body dynamics allows for a larger variation of evolutionary pathways when compared to binary systems. This includes mass transfer from the tertiary onto the inner binary \citep{zwart_triple_2019, gao_stellar_2022, dorozsmai_stellar_2024, kummer_in_prep}, triple common envelope \citep{glanz_simulations_2021}, and dynamical destabilization \citep{perets_triple_2012, hamers_return_2022, toonen_stellar_2022}, which can lead to either the ejection of a component or a collision between two bodies.

A triple system destabilizes when the ratio between the outer and inner semi-major axis decreases, which occurs due to mass loss in the inner binary from stellar winds \citep{perets_triple_2012}, mass transfer in the inner binary when the donor becomes less massive than the accretor \citep{hamers_return_2022}, supernova kicks in either the inner or outer binary, or from the gravitational perturbation of a passing object \citep{michaely_high_2020, hamers_return_2022, stegmann_close_2024}. The destabilization of triples due to mass loss is known as Triple Evolution Dynamical Instability (TEDI) \citep{perets_triple_2012}, and may be responsible for a large population of Galactic stellar collisions, in addition to being an efficient pathway for producing single stars through the ejection of one of the bodies. Though the potential outcomes of unstable triples are known from theory, a complete statistical overview of physical systems is lacking, with a few recent studied having provided more insight into these numbers. \cite{toonen_stellar_2022} (hereby referred to as TBZ22) performed N-body simulations of low-mass triple systems on the edge of stability, and find that up to $\sim 45$\% of these systems disintegrate into an ejected single star and a remaining binary, with the single star often reaching velocities of tens of km/s. Additionally, between $13$ \% and $24$ \% of the unstable triples produce a collision between two bodies, wherein the vast majority happens between two main-sequence stars. \cite{hamers_return_2022} (hereby referred to as HPTN22) performed a similar study of systems that undergo TEDI, and likewise found that unstable triples can produce stellar collisions and unbound stars at a Galactic rate of $10^{-4} \, {\rm yr}^{-1}$. In both of these studies, the mass of primary star was sampled from a Kroupa distribution \citep{kroupa_variation_2001}. TBZ22 excluded stars with masses greater than $7.5$ M$_\odot$, while HPTN22 sampled the full mass range. Consequently, a similar study has not yet been produced for triple systems with exclusively massive primaries. As already mentioned, the vast majority of massive stars are found in triples or higher order multiples, which means that a full consideration of triple dynamics is crucial in the context of understanding the evolutionary pathways of massive stars. Their shorter lifetimes and distinct orbits combined with strong stellar winds means that the orbits of massive multiples are likely to change on shorter timescales than their lower mass counterparts, potentially resulting in a higher fraction of triples that become unstable within a Hubble time. 

Unstable low-mass triples may lead to merger products and ejected single stars and binaries. If the same holds true for massive stars, triples may partly explain the large numbers of observed massive runaway stars \citep{gies_kinematical_1987, stone_space_1991, oey_resolved_2018, renzo_massive_2019, stoop_early_2023}, which can not be explained fully by supernovae in binaries or dynamic interactions in clusters \citep{de_wit_origin_2005, gvaramadze_field_2012}. Additionally, as massive stars end their lives as compact remnants - neutron star (NS) or black hole (BH) - massive triple systems that survive one or more supernova explosions may yet get pushed to the stability limit, after which they disintegrate and produce outcomes like runaway black holes/neutron stars, or merger/collisions between compact remnants or a compact remnant and a star, which may lead to phenomena such as tidal disruption events (TDEs) and gravitational wave sources.

In this study we follow up on TBZ22 by focusing exclusively on triples with a massive primary star. With a combination of numerical techniques for both secular orbit-averaged evolution and N-body dynamics, we follow the evolution of triples starting from a stable hierarchical configuration at the ZAMS, through the regime of dynamical instability, and until the potential disintegration of the system. In the following section we present a brief overview of three-body dynamics and the topic of stability in hierarchical triples (Sect. \ref{section:theory}), followed by an overview of the methodology in which we discuss the choices related to our numerical setup (Sect. \ref{section:methods}). We then present the results of our simulations and the statistical properties of the outcomes, including the components involved in collisions and the velocities of ejected stars (Sect. \ref{section:results}). Finally, we discuss our results in the context of observable phenomena (Sect. \ref{section:dicussion}), before presenting our conclusions as a summary of our results along with suggestions for future work (Sect. \ref{section:conclusion}).

\section{Three-body dynamics}\label{section:theory}

\subsection{Hierarchical triples}

As opposed to the case of two gravitationally bound objects, the addition of a third component drastically changes many aspects of the system, one of which is the stability. A two-body system that does not experience any external or internal changes will remain in its orbital configuration indefinitely, and the state vectors of the components can be calculated using analytical formulae. In a three-body system, no analytical solution exists and the dynamics must therefore be solved using numerical methods. Most three-body systems are chaotic, meaning their outcome is sensitive to the initial conditions, wherein the degree of sensitivity depends on the configuration of the system \citep{boekholt_gargantuan_2020, boekholt_gargantuan_2023}.

Triples in the universe have so far been found in a configuration consisting of an inner binary whose centre of mass is in a wide orbit with a tertiary companion. This configuration, known as a hierarchical triple, occurs when the separation between the components in the inner binary, denoted as $a\ssin$, is significantly smaller than that of the outer binary, denoted by $a\ssout$. The larger the ratio $a\ssout/a\ssin$, the stronger the hierarchy and the longer the system can remain stable. The hierarchical configuration allows the system to be described as two nested Keplerian orbits with semi-major axes $a\ssin$ and $a\ssout$, eccentricities $e\ssin$ and $e\ssout$, arguments of periapsis $g\ssin$ and $g\ssout$, and longitudes of ascending nodes $\Omega\ssin$ and $\Omega\ssout$. The two orbits can be inclined with respect to each other. This mutual inclination, denoted as $i$, as is defined as the angle between the orbital angular momentum vectors of the inner and outer orbits. The components in a hierarchical triple are denoted as the primary, secondary, and tertiary, with $m_1 \geq m_2$ labeling the masses of the primary and secondary, and $m_3$ the tertiary. 

A triple with a stronger hierarchy (higher value of $a\ssout/a\ssin$) can remain in its configuration for longer. However, stability is determined by a combination of orbital and stellar parameters, and not just the separations alone. There have been several proposed prescriptions for estimating the stability of hierarchical triple systems, one of which is the \cite{mardling_dynamics_1999} criterion, which provides an analytical estimate for the critical semi-major axis ratio, defined such that a triple is classified as unstable if the actual semi-major axis ratio  $\frac{a\ssout}{a\ssin}$ is smaller than the critical ratio $\frac{a\ssout}{a\ssin}\vert_{\rm crit}$, where

\begin{equation}\label{eq:mardling_aarseth}
    \frac{a\ssout}{a\ssin}\bigg\rvert_{\rm crit} = \frac{2.8}{1 - e\ssout} \left(1 - \frac{0.3i}{\pi} \right)\left(\frac{1 + q\ssout)(1 + e\ssout)}{\sqrt{1 - e\ssout}} \right)^{2/5}.
\end{equation}

\noindent Here, $q\ssout$ is the outer mass ratio defined as $q\ssout \equiv m_3/(m_1 + m_2)$.

\subsection{The Kozai-Lidov mechanism}

The von Zeipel-Lidov-Kozai (ZKL) \citep{lidov_evolution_1962, kozai_secular_1962} mechanism is one of the most important effects in a hierarchical triple, as it can drastically affect the evolution of the system and its stellar components. ZKL is a secular process that introduces periodic changes in the mutual inclination and inner eccentricity of a hierarchical triple system. The physical mechanism behind the process is described in detail in e.g. \cite{naoz_eccentric_2016}, but can be summarized as a torque between the two orbits, allowing them to exchange angular momentum while conserving energy. This causes the orbits to alter their shapes and orientations. The ZKL process is also known as the ZKL cycle, as it can cause the system to oscillate between high inclination and low inner eccentricity, to a low inclination and high eccentricity, in a periodic cycle.


\section{Methodology}\label{section:methods}

We combine two different methods for studying unstable triples: orbit-averaged dynamics with stellar evolution for long-timescale simulation of orbits and stellar objects, and a direct N-body approach. In the orbit-averaged technique, differential equations describing the time evolution of the orbital elements are solved, as opposed to the positions and velocities of the bodies. As these change on timescales significantly longer than the dynamical timescales, systems can be simulated for long periods at a fraction of the computational cost when compared to direct simulations. In this study, a population of massive triple system are evolved from zero-age main sequence using the triple population synthesis code \tres{} \citep{toonen_evolution_2016, kummer_main_2023}. The subset of systems that cross the stability limit (Equation \ref{eq:mardling_aarseth}) are then further evolved using a dynamical approach in an N-body code.

At the onset of instability, the orbit-averaged approach of secular evolution can no longer be utilized as this technique is only valid for strongly hierarchical systems. On the stability limit, perturbations on the inner orbit from the outer star may occur on timescales on the order of the dynamical timescales, meaning that these changes are not captured by the orbit-averaged approach. To further study the evolution of these now-unstable triple systems, they therefore have to be simulated dynamically using an N-body approach. 

\subsection{Evolving triples with \tres{}}

TRiple Evolution Simulation (\tres{}) is a code developed with the goal of rapid evolution of large populations of hierarchical triples. For a full overview of the internals of \tres{}, see \cite{toonen_evolution_2016}. In \tres{}, stellar evolution is included as single star evolution implemented using the fast stellar evolution code SeBa \citep{portegies_zwart_population_1996, toonen_supernova_2012}, which is based on fitted stellar evolution tracks from \cite{hurley_comprehensive_2000}. SeBa provides fast evaluations for stellar parameters as a function of the initial stellar mass and time. Evolution of the orbital parameters is implemented in \tres{} as a system of ordinary differential equations (ODEs) derived using time-independent perturbations of the Hamiltonian in the semi-major axis ratios \citep{ford_secular_2000}. The set of ODEs includes the inner and outer orbital parameters, in addition to total angular momentum, mutual inclination, and spin frequencies of each of the three stars. \tres{} includes both the lowest order expansion of the Hamiltonian, the quadrupole-level, and the octupole-level, in addition to gravitational wave emission, precession, and tidal friction. The code evolves triple systems until a specific stopping condition is reached. In the simulations studied in this work, the specific stopping conditions are:

\begin{itemize}
    \item Primary or secondary Roche lobe overflow.
    \item Tertiary Roche lobe overflow.
    \item System crosses the stability limit defined by the stability criterion in Equation \ref{eq:mardling_aarseth}.
    \item Inner orbit becomes unbound due to core-collapse supernova in either the secondary or primary.
    \item Outer orbit becomes unbound due to core-collapse supernova in any of the three components.
    \item The system reaches Hubble time without any of the aforementioned stopping criteria being triggered.
\end{itemize}

\noindent Here we present a summarized version of the choice of initial conditions for the triple population studied in this work, based on \cite{kummer_main_2023}. For a full description, see the aforementioned paper. The initial population of triples are generated by sampling distributions of the component masses and orbital elements. The generated systems consists of co-evolving stars at the zero age main sequence (ZAMS) bound together in an initially dynamically stable configuration. Given that these are systems containing massive stars, there are large uncertainties in the distribution of stellar and orbital parameters due to the low number of observed counterparts. To compensate for this, \cite{kummer_main_2023} produces sets of varied initial conditions by applying a series of assumptions. Specifically, they define their fiducial model which follows a Kroupa distribution \citep{kroupa_variation_2001} for the primary mass, while the secondary and tertiary masses follow from the inner and outer mass ratios $q\ssin \equiv m_1/m_2$ and $q\ssout \equiv m_3/(m_1 + m_2)$, which are both sampled from a flat distribution in $[0.1, 1]$\footnote{This distribution is based on observations of massive binaries from \citet{sana_binary_2012} and \citet{kobulnicky_toward_2014}.}. For the orbital parameters, the semi-major axes $a\ssin$ and $a\ssout$ are both sampled from various distributions such that systems that are initially overflowing their Roche Lobe at the ZAMS are discarded, in addition to systems that are far enough apart for them to be weakly gravitationally bound. This gives limits of $a_\text{in, out} \in [5, 5\times 10^6]$  R$_\odot$. The eccentricities $e\ssin$ and $e\ssout$ are both sampled from thermal distributions in the range $[0, 0.95]$ \citep{ambartsumian_statistics_1937, heggie_binary_1975, moe_mind_2017, hwang_mystery_2023}. The arguments of pericenter $g\ssin$ and $g\ssout$, along with the mutual inclination $i_{\rm mut}$ were sampled using uniform distributions in $[-\pi, \pi]$ and $[0, \pi]$ respectively. Finally, stellar rotational velocities were sampled using the distribution described in \citet{hurley_comprehensive_2000}.

\subsection{From orbit-averaged to dynamical evolution}

As the orbit-averaged approach utilized by \tres{} has no information about the positions and velocities of each body at any given time, and consequently, we do not have full initial conditions for the N-body integrator a priori. Therefore, when initializing a triple evolved with \tres{} in the dynamics code, we need to make a choice for the anomaly, i.e. the initial angular positions in the orbits at the start of the simulation. When a triple system is on the edge of a stability, the hierarchical configuration might be lost within a few outer orbits \citep{toonen_stellar_2022}, which means that the initial positions of the bodies in the system is likely to play a larger role in the time it takes for the system to disintegrate and consequently the final outcome. Therefore, randomly positioning each system in the orbit is likely to give us only a small sample of the potential outcomes of the population. To mitigate this, we make ten instances of each system, and distribute them according to a uniform distribution of the true anomaly $\theta \in [0, 2\pi]$, where an anomaly of $0$ corresponds to periapsis. In other words, each system has ten different variations, where the only difference is the initial true anomaly of the outer orbit. TBZ22 performed similar initializations of the systems in their study. The reason for varying the outer anomaly as opposed to the inner, is that it is the outer orbit that has the most effect on the stability of the system, as seen by the dependency on $e\ssout$ in the stability criterion (Eq. \ref{eq:mardling_aarseth}).

\subsection{Simulating dynamics with \texttt{Syzygy.jl}}\label{sec:dynamics_with_syzygy}

 For this study, we utilize the package \texttt{Syzygy.jl}\footnote{Code is available at \url{https://github.com/casparwb/Syzygy.jl}}, written by the main author in the \texttt{Julia} programming language\footnote{Official website: \url{https://julialang.org/}}. The package utilizes the \texttt{DifferentialEquations.jl} (\texttt{DiffEq}) \citep{rackauckas_differentialequationsjl_2017} ecosystem to solve the governing differential equations, providing a combination of high performance, stability, and flexibility, as it allows the users to choose from a large variety of optimized ODE solvers. In this study, we utilize the 8th order explicit adaptive Runge-Kutta-Nyström method with absolute and relative error tolerances of $10^{-10}$. We found that this combination of algorithm order and error tolerances provided the optimal balance of speed and long-term stability. We include a set of criteria to stop the simulations in the case of specific events. These stopping conditions are implemented as callbacks, which, in the \texttt{DiffEq} ecosystem, is defined as code that is injected by the user into the solver algorithms. Callbacks allow the solver to handle events, including discontinuities, in a safe and stable way, though in the case of stopping conditions they are simply functions that, at each time step, takes in the state of the integrator, checks one or more criteria and terminates the integration in the case of the criterion being fulfilled. In this study we implement four stopping conditions:

\begin{itemize}
    \item Collision between two bodies: if any two bodies have overlapping radii, we stop the simulation and flag the outcome as a collision with the two components. See section \ref{sec:discussion_collisions} for a discussion of this.
    \item Unbound: if a body gets ejected due to a close passage, we stop the simulation and flag the outcome as escaper. To check whether a body, at any point, has become unbound, we implement the constraints from \cite{standish_sufficient_1971}, which consists of three criteria that must be fulfilled: (1) the body is a given distance away from the centre of mass of the other two bodies. We set this distance to be $100$ times the oscullating semi-major axis of the remaining two bodies. (2) The body is moving away from the centre of mass of the remaining components, and (3), the body has positive energy. It may occur that the third criteria is not satisfied even when a body has been ejected far from the other components, meaning that a tiny background gravitational potential (as is present in any astrophysical gravitational system) would unbind the ejected star. As a consequence, we flag the body as a drifter if only the first two criteria are satisfied, and the body is a distance of 1 parsec from the remaining binary centre of mass.
    \item To set a limit on the amount of computational resources we use, and to avoid running the simulations for unpractical timescales, we stop the simulation if a system is simulated for longer than a preset maximum CPU time. For the simulations in this study, we set the max CPU time to be $10$ hours. 
    \item The system reaches the maximum allotted simulation time without any of the aforementioned conditions being triggered.
\end{itemize}

\texttt{Syzygy.jl} also includes flags for checking whether a system remains in a hierarchical configuration, or if it has entered a so-called democratic phase, wherein the hierarchy has broken down. These are not stopping conditions, but rather bookkeeping for later analysis. This is an important inclusion when analysing the systems that eject one of the components. There are two pathways that can cause a body to be ejected from the triple: (1) ZKL effects in triples close to the instability limit can cause extreme eccentricity excitations in the outer orbit, resulting in a small outer periastron distance where the tertiary can reach and exceed the escape velocity of the system due a nudge from the inner binary. (2) The system enters a democratic phase, in which close passages can occur between any pair of stars, potentially accelerating one of the components to escape velocities. For the systems that eject a component, we label the outcome as either a \textit{Hierarchical} or \textit{Democratic-X} ejection, depending on whether the democratic flag was raised at any point during the simulation, with X being either 1, 2, or 3, denoting which component was ejected. We implement three different methods for checking whether the triple remains hierarchical: 

\begin{enumerate}
    \item Check whether the pair of bodies with the smallest separation is the initial inner binary. This test is quick but it might be prone to false positives in systems with highly eccentric outer orbits.
    \item Check if the original inner binary is still the binary with the smallest semi-major axis. This check is more computationally expensive, as it requires the calculations of the total energy in each pair to check if they are bound, followed by the calculation of the semi-major axis. It is, however, more robust than the first check.
    \item Check if any of the components in the inner binary have hyperbolic orbits ($e \geq 1$). This check should in theory only be true if the system is not hierarchical, however the osculating orbital elements might be prone to instantaneous deviations from the expected values due to numerical inaccuracies.
\end{enumerate}

Additionally, we implement the ability to raise flags if a body overflows its Roche lobe (RL). For the tertiary component in a hierarchical triple, its RL is calculated by approximating the system as a binary, where the companion mass is equal to the total mass of the inner binary. However, unstable systems can lose their hierarchy \citep{toonen_stellar_2022}, which begs the question of how to check for Roche lobe overflow (RLOF) if the system is in a democratic phase. To implement this, we include two callbacks to check for RLOF, one for while the system is hierarchical, and one for if the hierarchy has broken down. We use the previously mentioned criteria to check if the democratic flag has been raised, and then use the appropriate check to test for RLOF. For a system in which the democratic flag is raised, we go through each body in the system, find the nearest other body, and use the generalized volume-equivalent Roche Lobe radius \citep{eggleton_aproximations_1983, sepinsky_equipotential_2007}:

\begin{equation}
    R^{\rm Egg}_L(t) = D(t) \frac{0.49q^{2/3}}{0.6q^{2/3} + \ln{(1 +q^{1/3})}},
\end{equation}

where $D(t)$ is the distance between the two bodies at a time $t$, and $q$ is their mass ratio. If the radius of a star is ever greater than $R^{\rm Egg}_L(t)$, we store the time and the identities of the mass-transferring components. We emphasise that the RLOF check is simply a flag, and not a stopping condition. Realistically, mass transfer during RLOF would alter the dynamics and likely change the outcome of these simulation. However, eccentric mass transfer is still poorly understood (see \cite{sepinsky_equipotential_2007,church_mass_2009,hamers_analytic_2019} \cite{dosopoulou_roche-lobe_2017}), and prescriptions suitable for N-body simulations are non-existent. Therefore, we simply flag these systems to get an overview of the occurrence of RLOF in unstable massive triple systems, and make a note that these systems might have a different outcome if mass transfer was included.

\subsubsection{Stellar evolution}

In this study we simulate the triple population using two variations of dynamical evolution: a pure N-body approach in which only Newtonian gravity is included, and an N-body + stellar evolution model. The latter incorporates the effects of stellar evolution through changes in mass and radius. As showed in TBZ22, triple stars that reside on the stability limit may remain hierarchical for thousands and up to tens of millions of crossing times\footnote{TBZ22 defines the crossing time as $t_{\rm cross} \equiv R_v/V$, where $R_v = (GM_{\rm tot}^2)/(-2E_{\rm tot})$ is the characteristic size, and $V = \sqrt{(-2E_{\rm tot}/M_{\rm tot}})$ is the velocity, with $M_{\rm tot}$ being the total mass of the system and $E_{\rm tot}$ the total energy.\label{crossing_time_fn}}, which means that some systems will need to be simulated for timescales approaching or exceeding the nuclear timescales before destabilization occurs. As a consequence, the effect of stellar evolution should be taken into account to get a more realistic view of the dynamics of the system. Since this study is looking at massive triples, stellar evolution is likely to play an even bigger role in the dynamics due to the substantially smaller nuclear timescales and higher mass loss when compared to their low-mass counterparts. 

In order to incorporate stellar evolution into the direct simulation without adding substantial computational cost, we use \texttt{SeBa} to pre-calculate the single stellar evolution for each component in each system. The system is evolved until every component becomes a remnant (white dwarf, neutron star, or black hole). We implement stellar evolution into the gravity simulation by constructing an interpolator for each stellar parameter, followed by setting up a callback which takes in the current state of the integrator and interpolates the parameters at each time step. The masses, radii, and stellar types of each object are then updated with the interpolated values. Using a callback allows for the ODE system to remain stable in the case of sudden mass loss due to e.g. a supernova (SN). We also include instantaneous velocity kicks on the bodies that experience supernovae using the SN kick model of \cite{verbunt_observed_2017}, combined with a fallback model \citep{fryer_compact_2012} to reduce the kick velocities of higher mass black holes.

\section{Results}\label{section:results}
\subsection{Secular evolution}

\begin{figure}
    \centering
    \includegraphics[width=0.45\textwidth]{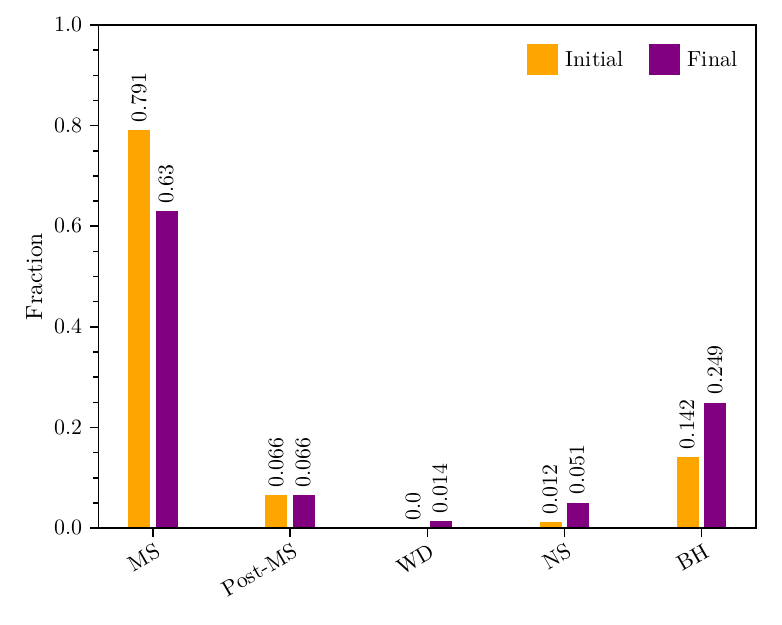}
    \caption{Distributions of stellar types for the systems that become dynamically unstable from secular evolution. The purple bars indicate the stellar types at the end of the simulations where SE is included.}
    \label{fig:stellar_types_distribution}
\end{figure}

In a typical population of massive triples, with minimum primary masses of $10$ M$_\odot$, there are several potential evolutionary channels that the systems may evolve through. As shown by \cite{kummer_main_2023}, $67.3$\% of the systems undergo mass transfer in the inner binary, $29$\% experience unbinding in either the inner or outer orbit due to a supernova, $1.8$\% undergo mass transfer from the tertiary onto the inner binary, while $1.7$\% become dynamically unstable due to winds in the inner binary or supernovae. At the onset of instability, the triple population contains $71$\% main sequence stars and $26$\% black holes, with the remaining percentages consisting of post-main sequence stars and neutron stars. In the case of our simulations with only gravity, this distribution of stellar types remains fixed, while the inclusion of SE means that the stars are evolving as they are simulated with the N-body code, resulting in a final distribution of stellar types that is distinct from the original (Fig. \ref{fig:stellar_types_distribution}).

\subsection{Duration of dynamically unstable phase}

\begin{figure*}
    \centering
    \begin{subfigure}{0.45\textwidth}
        \includegraphics[width=\textwidth]{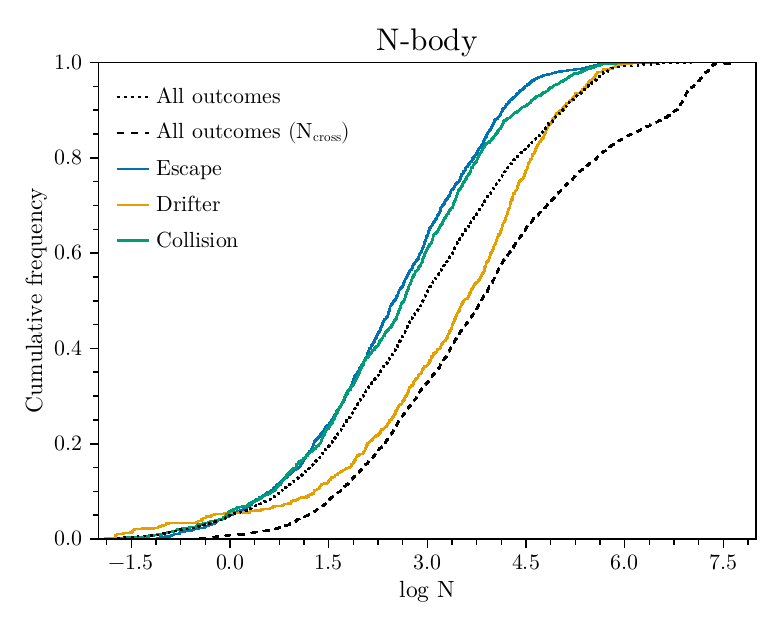}
        \caption{}
        \label{fig:duration_nbody}
    \end{subfigure}
    \hspace{1cm}
    \begin{subfigure}{0.45\textwidth}
        \includegraphics[width=\textwidth]{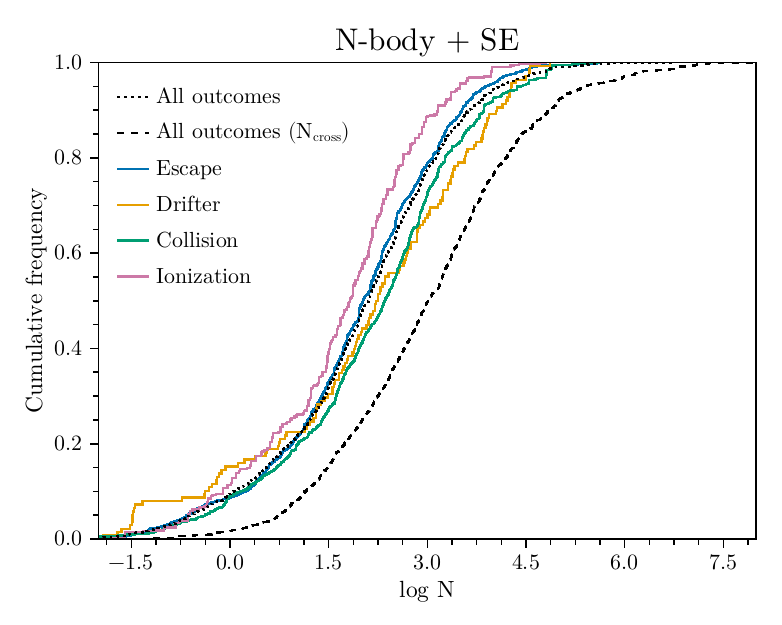}
        \caption{}
        \label{fig:duration_SE}
    \end{subfigure}
\caption{Duration of the unstable phase for the most prominent outcomes. The x-axis shows the total simulation time expressed as a multiple of the initial outer orbit, denoted with $N$. The dotted line shows the distribution of $N$ for all outcomes, while the dashed line shows the duration in units of crossing times\footref{crossing_time_fn}. Subfigures (a) and (b) show $N$ for the pure gravity and SE simulations respectively.}
\label{fig:duration_of_unstable_phase}
\end{figure*}

The maximum simulation time for triples is set to $10^7\times P_\text{out, init}$, where $P_\text{out, init}$ is the period of the outer orbit at the time of destabilization. However, the majority of the simulations were terminated before reaching this time due to another stopping condition being triggered. Figure (\ref{fig:duration_of_unstable_phase}) shows cumulative histograms of the simulation time divided by the initial outer period ($N \equiv t_{\rm end}/P_{\rm out, init}$). Half of all the systems have not triggered any stopping condition after $N=1000$, and $20\%$ of systems remain stable after a simulation time of more than $N=30 000$. These results show that triples are likely to remain hierarchical for long timescales despite being classified as unstable, an observation that coincides with that of TBZ23. 


\subsection{All outcomes}

\begin{figure}
    \centering
    \includegraphics[width=0.45\textwidth]{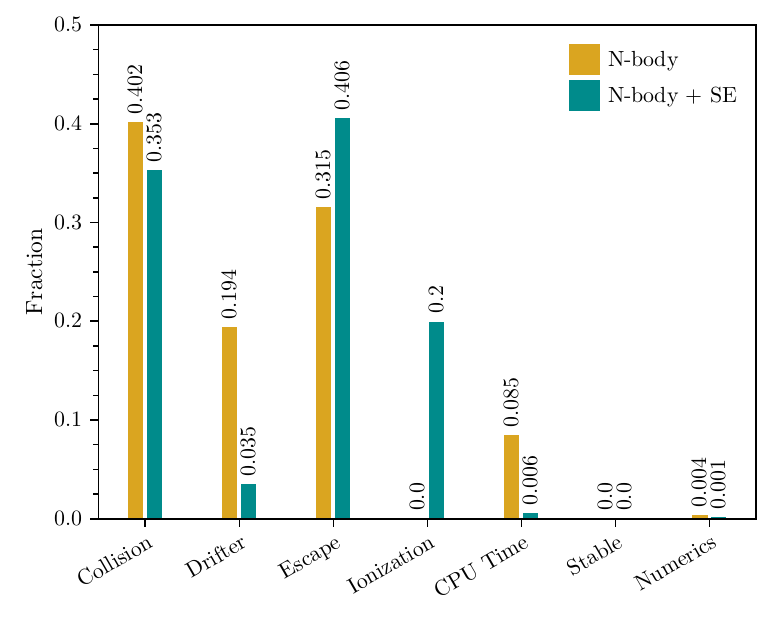}
    \caption{Frequencies of the different outcomes of the dynamical simulations for both the pure N-body and the SE models. The maximum simulation time was $10^7 \times P_{\rm out, init}$.}
    \label{fig:all_outcomes}
\end{figure}

For the simulations in which only gravity is taken into account, we find that the systems have different potential outcomes; collision, drifter, escape, CPU time, and numerical issues (labeled as "Numerics" in Figure \ref{fig:all_outcomes}). When SE is taken into account, we find that the systems experience the same outcomes with the addition of ionisation due to supernova kicks (Fig. \ref{fig:all_outcomes}). 
\begin{itemize}
    \item We find that $40$\% of systems experience a collision between two bodies when only gravity is considered. This percentage is slightly lower when we include SE, at $35$\%. 
    \item For the system that produce an escaper, $31$\% eject a body in the pure gravity case, while this rate increases to $40$\% when SE is also taken into account.
    \item In the case of the drifters, we find a large discrepancy between the two simulation suites, with $19$\% of the systems experiencing this outcome when only gravity is involved, while the inclusion of SE reduces this to $3.5$\% of systems. 
    \item Ionisation, in which all three stars become unbound from each other, only occurs when SE is considered and when one or more supernovae take place. We find that $20$\% of triples undergo ionisation with the inclusion of SE.
    
\end{itemize}


We also find that only $0.6\%$ of triples reach the CPU time limit when we include SE, compared to only $8.5\%$ in the case of pure gravity. This is likely due to the inclusion of stellar winds. If a system has a large semi-major axis ratio, $a\ssout/a\ssin$, while still being classified as unstable (due a to high outer eccentricity, significant mass ratio, or a mutual inclination close to $90^\circ$), the simulation will be limited by the small timestep in the inner orbit. Consequently, the code needs a substantial amount of CPU time to complete an outer orbit. In the N-body model, a system in this configuration will remain so until it disintegrates, which, due to the small timestep, might take longer than the allotted CPU time. With the inclusion of stellar winds in the SE model, the inner binary may widen due to mass loss, which decreases the semi-major axis ratio and increases the time step used by the ODE solver. These systems will get pushed further into the instability region, and will therefore not only be faster to simulate, but are also likely disintegrate more rapidly. 

\begin{figure*}
    \centering
    \includegraphics[width=0.9\textwidth]{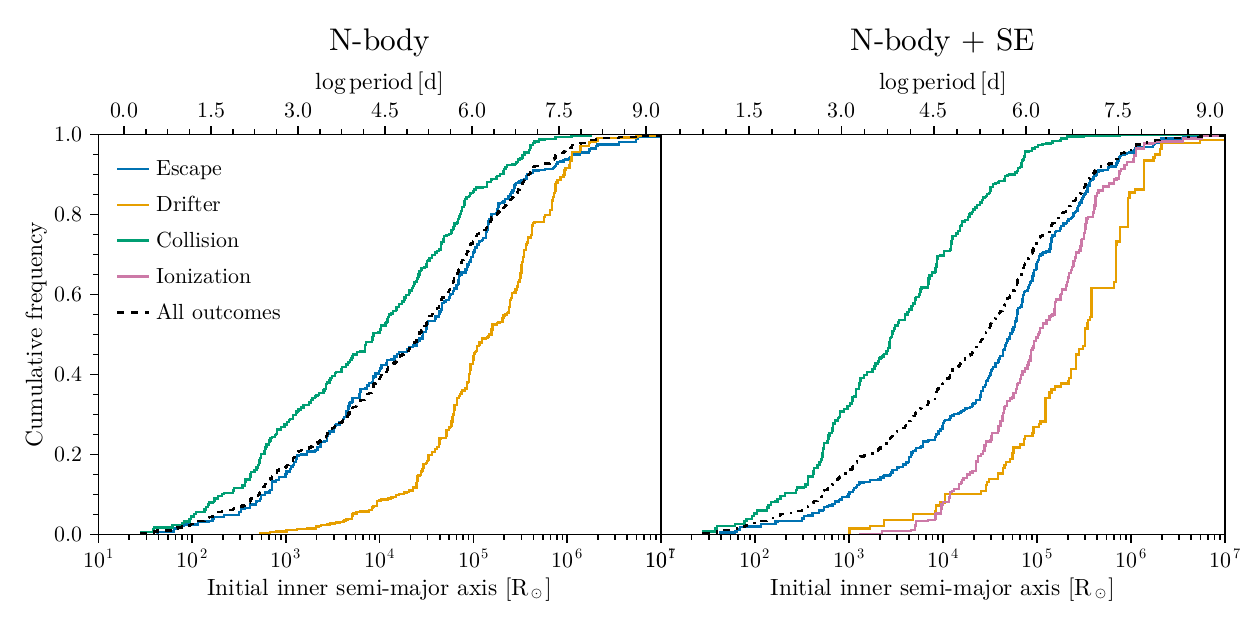}
    \caption{Cumulative histogram of the inner orbital separation at the onset of instability, color-coded by the final outcome of the system. }
    \label{fig:initial_sma_ecc_dist}
\end{figure*}

\subsection{Collisions}

\begin{figure}
    \centering
    \includegraphics[width=0.45\textwidth]{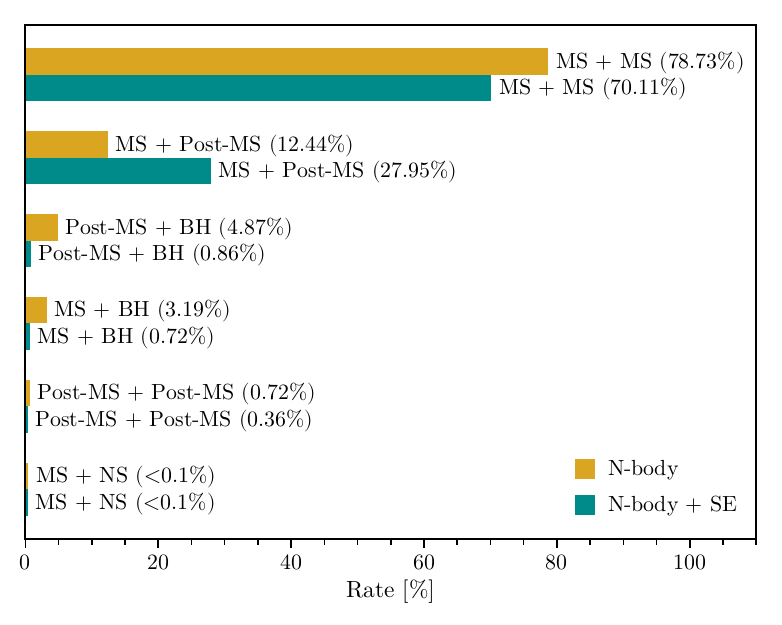}
    \caption{Percentages of the stellar types involved in collisions.}
    \label{fig:collider_types}
\end{figure}

Between $35\%$ and $40\%$ of the destabilizing systems experience a collision between two bodies. We find that the collisions happen between the primary and the secondary components in $96-98\%$ of the colliding systems, showing significant difference between the pure gravity systems and the systems with SE included. The stellar types of the collision pairs vary between the two models with the majority of collisions in both models taking place between two main sequence stars (Figure \ref{fig:collider_types}). Approximately $12.5 - 28$ \% of collisions occur between a MS and a post-MS star, with the larger percentage occurring when SE is not taken into account. When only gravity is included, we find that about $5$ \% and $3.2$ \% of collisions happen between (post-MS + BH) and (MS + BH) respectively, while these percentages are reduced to $0.86$ and $0.72$ for the SE model. 

\begin{figure*}
    \centering
    \includegraphics[width=0.9\textwidth]{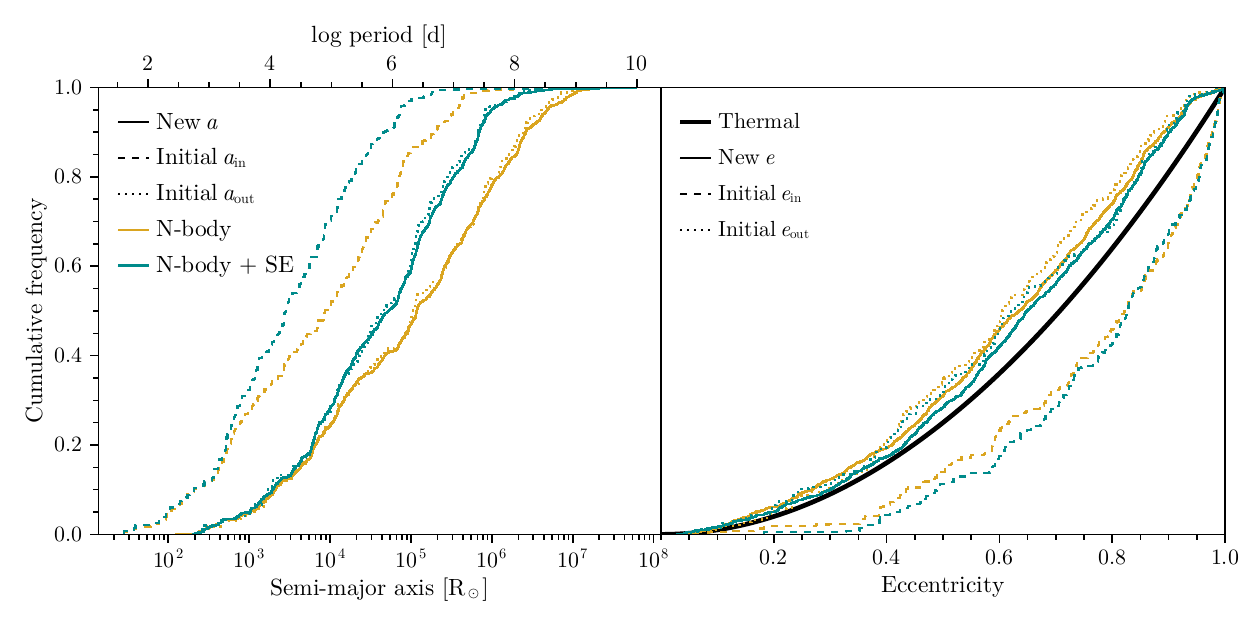}
    \caption{Properties of the newly formed binaries following the collision between two bodies in the original triple. The left figure shows the cumulative distribution of the semi-major axis, while the right shows the eccentricity.}
    \label{fig:properties_after_collision}
\end{figure*}

Following the collision, we calculate the properties of the newly formed binary using assumptions of a completely inelastic collision with mass conservation. We find that both simulation suites produce binaries with semi-major axes similar to the outer semi-major axis at the onset of instability. (Fig. \ref{fig:properties_after_collision}). 
\newline
\indent We observe a similar trend in the eccentricities of the newly formed binaries, namely that they tend to have eccentricities similar to the initial outer eccentricity of the triple progenitor. Given that the vast majority of collisions occur between the primary and secondary components, this result is to be expected. The new eccentricity distributions are also found to be sub-thermal, which differs from the results of binary-single scattering experiments \citep{heggie_binary_1975} by producing binaries that are generally less eccentric. A thermal eccentricity distribution is often used as the initial conditions of binary systems in population synthesis calculations, such as in SeBa. These results show that binaries formed from disintegrating triples are likely to deviate from this distribution.

\subsection{Ejections}

\begin{figure}
    \centering
    \includegraphics[width=0.45\textwidth]{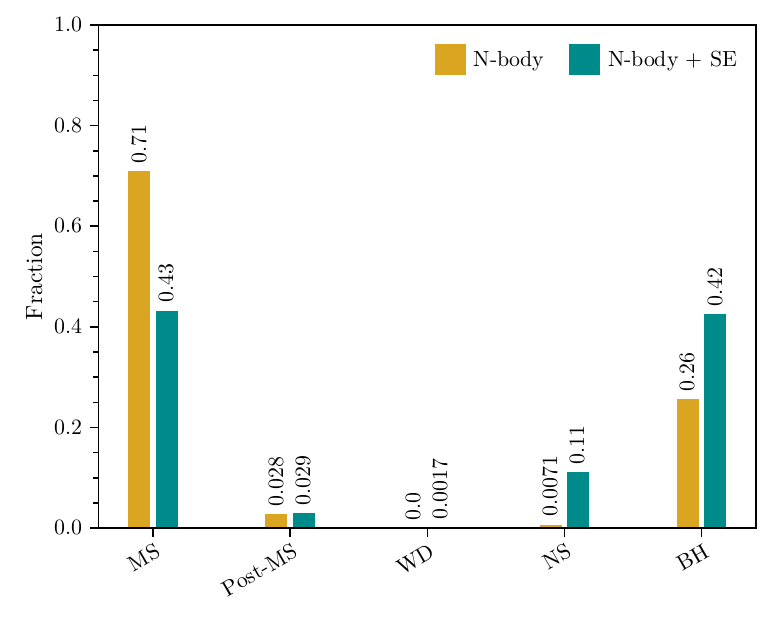}
    \caption{Distributions of the different stellar types that are ejected from the system during the unstable phase.}
    \label{fig:escaper_types}
\end{figure}

Between $31 \%$ and $41\%$ of the unstable systems result in the ejection of one of the components. When only gravity is included, the majority of the ejected bodies are main-sequence stars and black holes, with percentages of $71\%$ and $26\%$ respectively (Fig. \ref{fig:escaper_types}). When taking SE into account, these percentages are close to being equal, at $43\%$ and $42\%$ respectively, with the remaining escapers consisting mainly of neutron stars. 


\begin{figure*}
    \centering
    \includegraphics[width=0.9\textwidth]{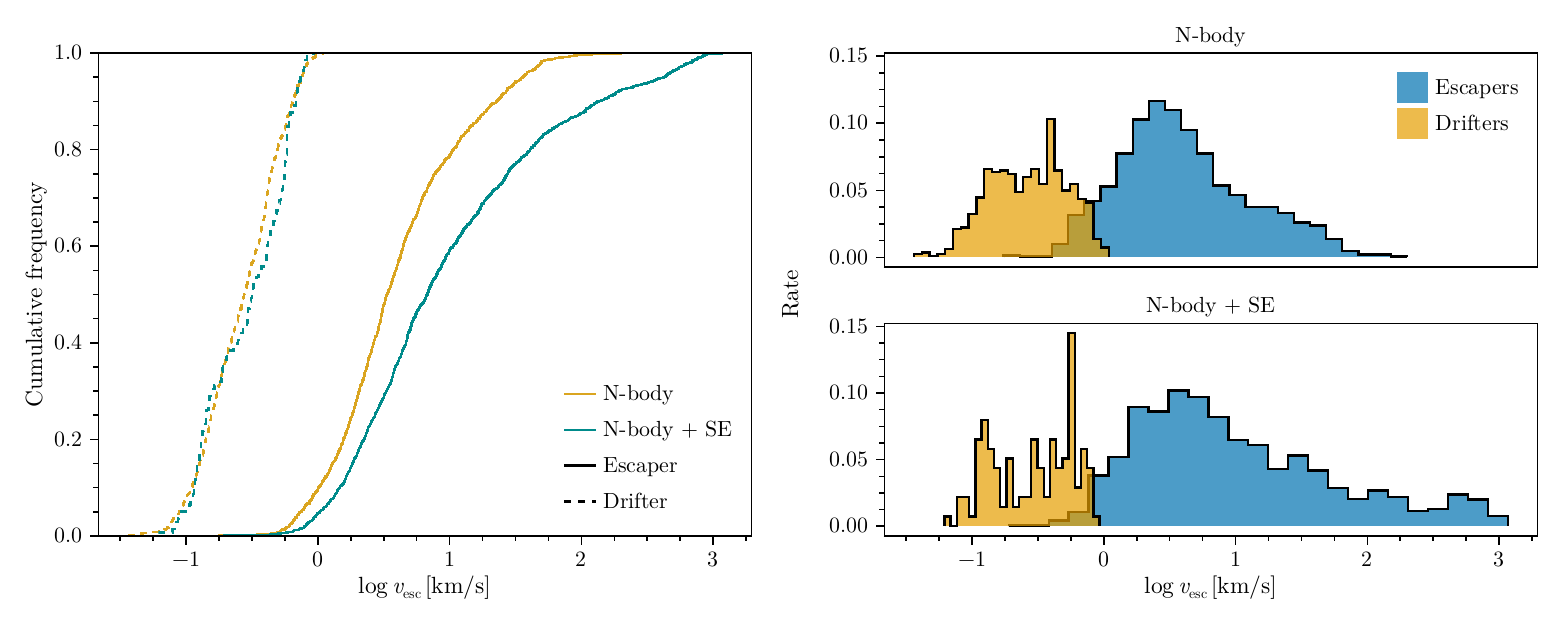}
    \caption{Velocities of the components labeled as drifters (orange) or escapers (blue). The same distribution is visualized as a cumulative frequency plot (left) and two histograms for each model (right).}
    \label{fig:runaway_velocities}
\end{figure*}

\begin{figure}
    \centering
    \includegraphics[width=0.45\textwidth]{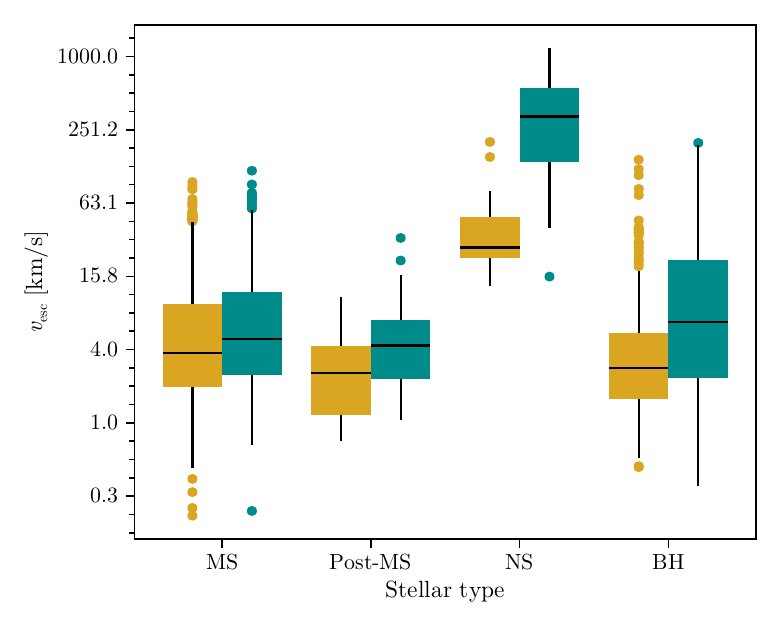}
    \caption{Distributions of the velocities of the ejected components as a function of their stellar types. The crossbar marks the median value of each distribution, with the whiskers indicating $1$ interquartile range. The markers show the outliers laying outside the whiskers.}
    \label{fig:escape_velocity_aof_stellar_type}
\end{figure}

Ejected bodies are categorized as either escapers or drifters depending on the criteria as explained in Section (\ref{sec:dynamics_with_syzygy}). One of the major differing characteristics of these bodies are their velocities (Fig. \ref{fig:runaway_velocities}). In both simulation suites we find that drifters have consistently lower velocities than their escaper counterparts, with a mean velocity of $0.3$ km/s. This is to be expected from the definition of the drifters: bodies that have been ejected to substantially large distances away from the remaining binary, but are still loosely bound to the system. When it comes to the escapers, the velocities differ more between the two models, with more high-velocity escapers being produced when SE is taken into account. The tail of higher velocity escapers in the SE model arise from the inclusion of supernova kicks in the simulations, and consequently the highest velocity objects are exclusively neutron stars. However, as seen in Figure (\ref{fig:escape_velocity_aof_stellar_type}), ejected neutron stars in the pure N-body model still reach velocities on the order of tens of kilometers per second without the help of a supernova kick. For the main sequence stars, the median velocity is $3.6$ km/s, with a standard deviation of $12.5$ km/s, which shows that dynamical interactions can produce single stars with a wide range of velocities. 
 
\begin{figure}
    \centering
    \includegraphics[width=0.45\textwidth]{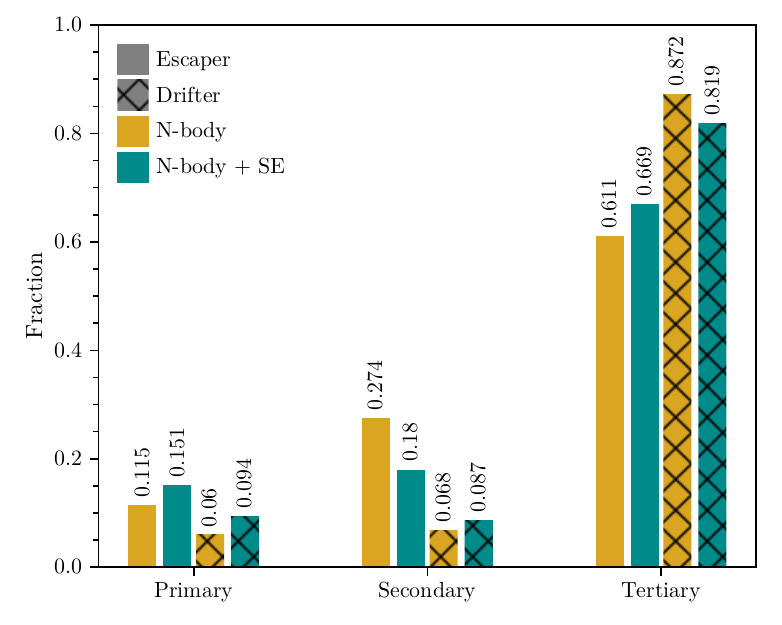}
    \caption{Histogram showing the frequency of which components are ejected. The solid and patterned bars indicate the distributions for escapers and drifters respectively, with the colors indicating the two different models.}
    \label{fig:which_component_escapes}
\end{figure}

We find that the original tertiary is the component that gets ejected in the majority of simulations. Without the inclusion of SE, the tertiary is ejected in $61$ \% of cases, while this percentage increases to $67$ \% when SE is taken into account (Fig. \ref{fig:which_component_escapes}). In the case of pure gravity, we observe about half as many primary ejections versus secondary ejections, while this difference is only $3$ \% when we include SE. When we look at the triples that produce drifters, we find that it is the tertiary that becomes a drifter in $82 - 87$ \% of these case, with the smaller percentage coming from the SE model. We expect that the tertiary is more likely to become a drifter due to the already lower binding energy in the outer orbit.

\begin{figure*}
    \centering
    \includegraphics[width=0.9\textwidth]{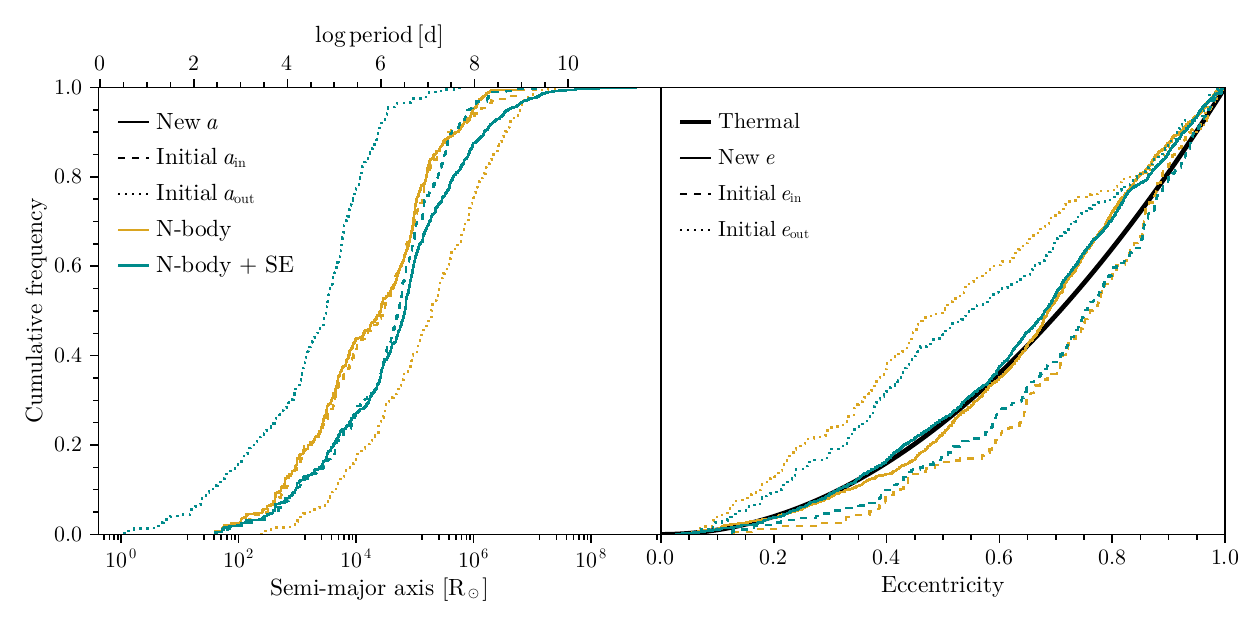}
    \caption{Properties of remaining binary following the ejection of one of the components.}
    \label{fig:properties_after_escape}
\end{figure*}

\begin{figure}
    \centering
    \includegraphics[width=0.45\textwidth]{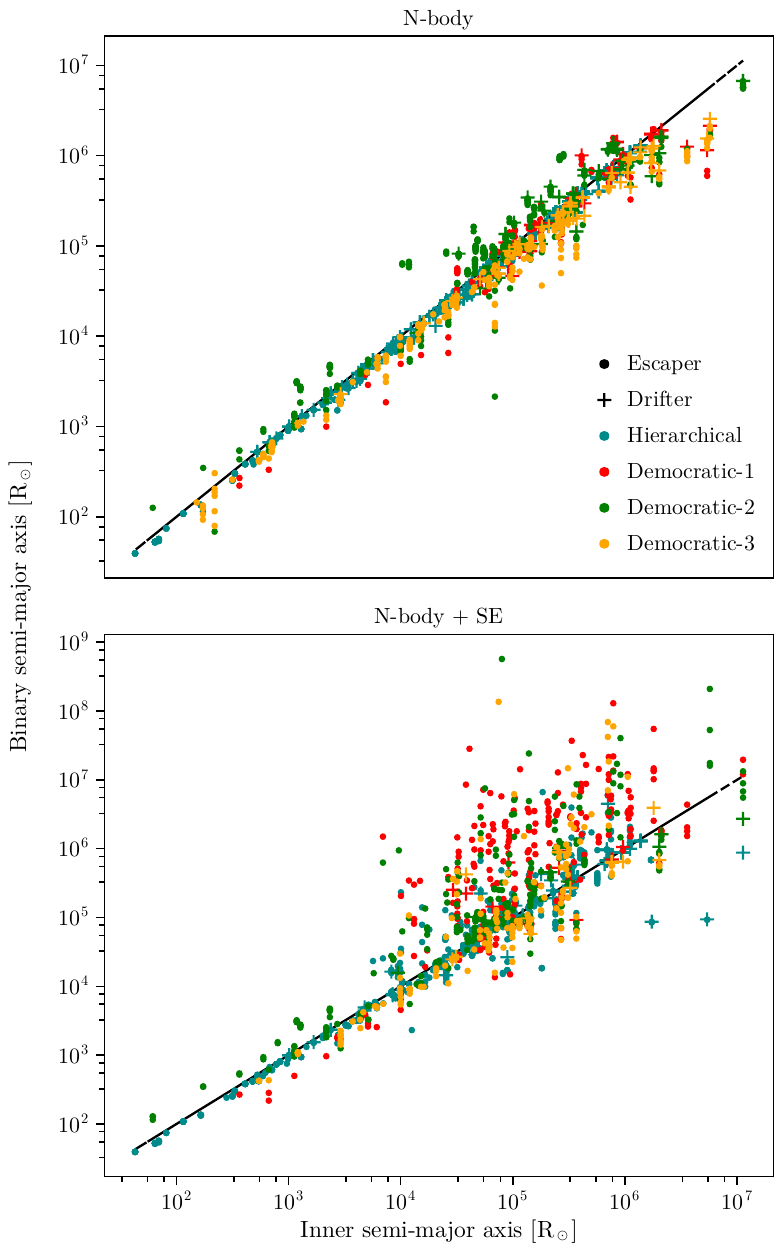}
    \caption{How does the semi-major axis of the newly formed binary following an ejection differ from the initial inner semi-major axis? The figure shows the change in separation for the N-body model (top) and the SE model (bottom), with different markers indicating whether the system produced an escaper or drifters, and the colors denoting which component was ejected. The solid line indicates a binary semi-major axis equal to the initial inner semi-major axis.}
    \label{fig:change_in_orbital_separation}
\end{figure}

The binaries that are formed following the ejection of one of the components tend to have separations similar to that of the inner semi-major axis at the onset of instability (Fig. \ref{fig:properties_after_escape}). This trend is strongest in the case of pure gravity, with the inclusion of SE producing more binaries with semi-major axes larger than the initial inner separation. This result is clearly visible in Figure (\ref{fig:change_in_orbital_separation}), which also shows how the new, wider binaries are mainly formed from democratic encounters where one of the stars of the inner binary is ejected. The hierarchical interactions, in which the tertiary is ejected without the democratic flag having been raised, is expected to produce new binaries with similar properties as that of the initially inner. We also find that both models produce binaries with eccentricities closely following a thermal distribution for $e \leq 0.6$, while deviating towards a sub-thermal distribution for larger eccentricities. 

\begin{figure}
    \centering
    \includegraphics[width=0.45\textwidth]{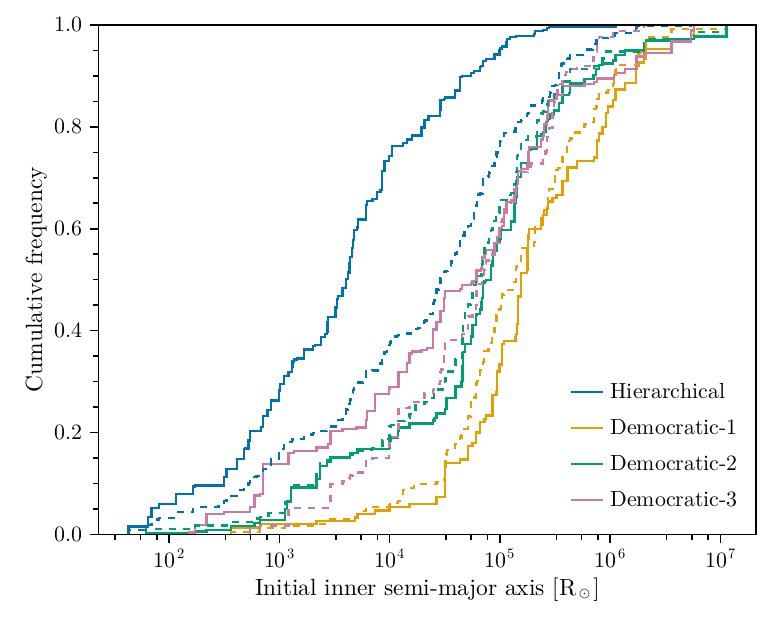}
    \caption{For the systems that eject one of the components, what are the distributions of the initial inner semi-major axis? This figure shows the cumulative frequency of this quantity for both the N-body model (solid) and the SE model (dashed). The different colours indicate whether the democratic flag was raised during the simulation, and which body was ejected. In the case of hierarchical ejection (blue lines), the tertiary was ejected without the democratic flag being raised, while the other colours indicate that the primary (yellow), secondary (green), or tertiary (magenta) was ejected after the flag was raised.}
    \label{fig:inner_sma_dependence}
\end{figure}

The value of $a\ssin$ at the onset of instability is correlated with the type of ejection. In the case of pure gravity, encounters that eject the tertiary without first having a democratic interaction tend to have smaller initial inner separations compared to the other types of ejections (Fig. \ref{fig:inner_sma_dependence}). When SE is considered, we find that the same category of ejection occurs in systems with a substantially higher $a\ssin$. 

\begin{figure}
    \centering
    \includegraphics[width=0.45\textwidth]{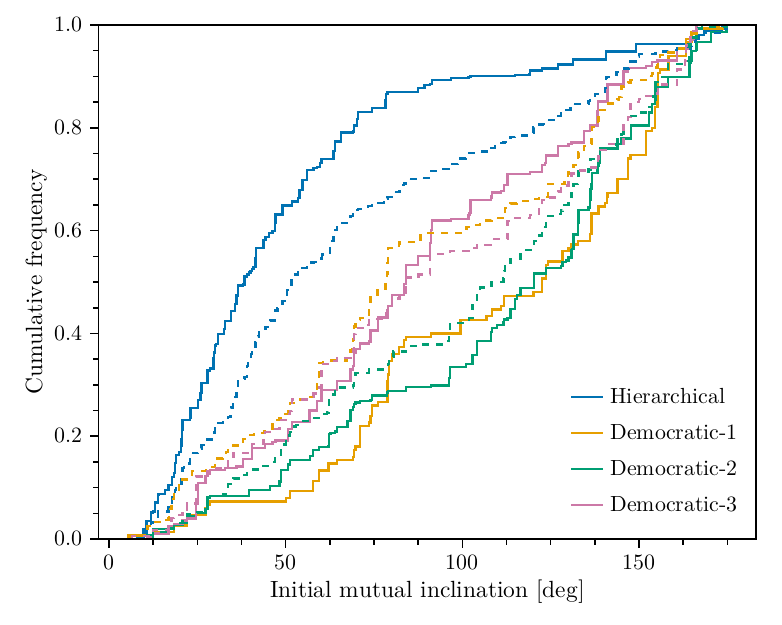}
    \caption{Similar plot as Figure (\ref{fig:inner_sma_dependence}), this time showing the cumulative distribution of the initial mutual inclination.}
    \label{fig:inclination_dependence}
\end{figure}

The initial mutual inclination also has a significant impact on whether the democratic flag is raised before the ejection of a body. The majority of hierarchical ejections in the case of pure gravity occur in systems with prograde orbits ($i \leq 90^\circ$), while non-hierarchical ejections are more evenly distributed over the full range of angles (Fig. \ref{fig:inclination_dependence}). When we include SE, the number of systems that eject a body without raising the democratic flag increases. We find that smaller initial inclinations tend to eject the primary (Democratic-1) when SE is included. Retrograde orbits ($i \geq 90^\circ$) are generally more stable than prograde orbits, but it is possible that the mass loss present in the systems with SE push the triples further into the instability regime where the effect of prograde versus retrograde orbits becomes less important.

\subsubsection{Mass of the ejected body}

\begin{figure}
    \centering
    \includegraphics[width=0.45\textwidth]{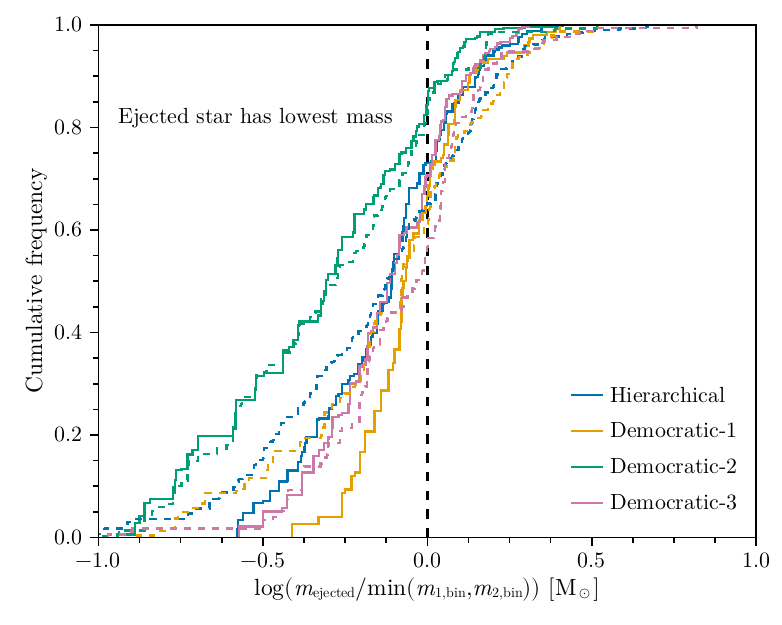}
    \caption{How often is the ejected star the most massive component? The figure shows a cumulative histogram of the mass of the ejected star over the smallest mass of the remaining binary. The different colours indicate the different ejection channels, with the solid lines corresponding to the case of pure gravity, and dashed indicating the inclusion of SE. The vertical, black dashed line indicates a value of 1, so that the left side of the plot corresponds to the ejected body having the lowest mass.}
    \label{fig:ejected_mass_over_minimum_binary_mass}
\end{figure}

In the systems that eject one of the components, the ejected body is the least massive in about $60 - 90$ \% of cases (Fig. \ref{fig:ejected_mass_over_minimum_binary_mass}), with the higher percentage representing ejections of the secondary (Democratic-2). This is similar to results found by TBZ22, who also note that one may naively assume the least massive body will always be the one to get ejected \footnote{This expectation comes from the binary-single star scattering experiments performed by \citet{hills_encounters_1975}, which showed that the least massive object is almost always the one that gets ejected.}. While this assumption indeed holds true for the systems that eject the secondary, it breaks down when we look at the other ejection channels, wherein the least massive body is ejected in around $65$ \% of cases.  

\section{Discussion}\label{section:dicussion}

\subsection{Instability rates}\label{sec:instability_rates}

As noted by HPTN22, the TEDI is triggered not only by mass loss, but also RLOF in the inner binary. Mass transfer is not included in \tres{} due to a lack of robust prescriptions for eccentric mass transfer. Therefore, our fractions for triples that become unstable is a lower limit. Including mass transfer in the models can potentially increase the rate of unstable triples by up to a factor 2, as seen in HPTN22. Furthermore, including perturbations caused by close fly-bys of other stars can also push triples over the stability limit, potentially increasing the fraction of TEDI systems by another percent. From \cite{kummer_main_2023}, almost $70$ \% of massive triples undergo mass transfer in the inner binary. If 1\% of these lead to an expansion of the inner separation, pushing the system over the stability limit, the fraction of unstable triples is increased by over $40\%$. As a consequence, we note that the rates calculated for the different outcomes should be considered a lower limit of the real population.

\subsection{Galactic rates}

To compute calactic rates for the different outcomes, we assume a Galactic star formation rate of 3 ${\rm M_\odot}/{\rm yr}$, a triple star fraction of $0.6$ \citep{offner_origin_2023}. The rates for different outcomes are shown in Table (\ref{tab:galactic_rates}). We see that the rates are all in the order of 1 event per Myr, with the largest rate originating from collisions in the pure N-body simulation, at 1.3 events/Myr. These rates are dominated by the initial mass function; only about $0.45$\% of all stars have masses $\geq 10 {\rm M_\odot}$ \citep{kroupa_variation_2001}. The Galactic rate for the destabilization of a triple due to stellar winds is estimated to be $3.4$ events/${\rm Myr}$. As already mentioned, this is likely a lower limit due the absence of mass transfer prescriptions in \texttt{TRES} and perturbations from fly-bys. If we assume that these mechanisms could bump up the fraction of systems that become unstable to $5$\%, as seen in HPTN22, the corresponding Galactic rate would then be 1 event per $10^5$ yr.


\begin{table}[]
\centering
\begin{tabular}{@{}lll@{}}
\toprule
Outcome                         & \multicolumn{2}{c}{Rate [${ \rm Myr}^{-1}$]} \\ \midrule
\multicolumn{1}{l|}{}           & \multicolumn{1}{l|}{N-body}        & N-body + SE        \\ \midrule
\multicolumn{1}{l|}{Collision}  & \multicolumn{1}{l|}{1.35}          & 1.13               \\
\multicolumn{1}{l|}{Escape}     & \multicolumn{1}{l|}{1.05}          & 1.30               \\
\multicolumn{1}{l|}{Drifter}    & \multicolumn{1}{l|}{0.65}          & 0.11               \\
\multicolumn{1}{l|}{Ionization} & \multicolumn{1}{l|}{0.0}           & 0.64               \\ \midrule
Destabilization                 & \multicolumn{2}{c}{3.36}                            
\end{tabular}
\caption{Estimated Galactic rates for the various outcomes of the unstable triples.}
\label{tab:galactic_rates}
\end{table}


\subsection{Ejections and runaways}

A large fraction of observed massive stars are moving with a high velocity relative to their environment. These stars make up about 5-10\% of all observed B-stars, and 10-30\% of observed O-stars \citep{gies_kinematical_1987, stone_space_1991, oey_resolved_2018, renzo_massive_2019, stoop_early_2023}, and were dubbed `runaway` stars by \cite{blaauw_origin_1961}. These massive runaway stars can have velocities up to 200 km/s, with a dispersion of 30 km/s \citep{stone_space_1991}. Historically, multiple scenarios have been proposed as mechanisms for producing massive runaways, with two channels being generally regarded as the most dominant; the disruption of a binary due to an internal supernova \citep{blaauw_origin_1961}, and the ejection of a star due to binary-single star interactions in dense stellar clusters \citep{fujii_origin_2011}. TBZ23 proposed a new channel for the formation of runaway stars; the destabilization of a triple system and the subsequent ejection of one of the components. They find that for low-mass stars, the destabilization of a triple system often leads to the ejection of one of the stellar components, with a velocity distribution of the ejected stars peaking at velocities of a few km/s, and including a tail up to orders of $10^2$ km/s. Additionally, the authors propose estimates of the maximum terminal velocity of ejected massive stars with mass $M=10 \, \rm M_\odot$ and a minimum distance of $10 \, \rm R_\odot$ to be $95 - 222$ km/s. Our simulations produce similar results, with a median velocity of $4.25$ km/s, and a tail that extends up to a maximum velocity of $116$ km/s. With our Galactic model, we estimate rates of $1 - 1.3$ ejections per Myr due to the destabilization of a massive triple. Out of all the ejected bodies, about $8$ \% of them have velocities $\geq 30$ km/s, and around half of these have masses $\geq 10$ M$_\odot$, which leads to a TEDI-induced massive, high-velocity runaway rate of about $4 - 5.4 \times 10^{-8}$ Myr$^{-1}$. Looking at percentages alone, by assuming a triple rate of $60$\% and a destabilization rate of $2$\%, combined with the results in this work, we find that about $0.02$\% of all massive stars become high-velocity runaways due to triple destabilization. If we assume that $20$ \% of all massive stars are runaways, then about $0.1$ \% of these originate from triple destabilization. Even with the assumption that this is a lower limit on the number of massive triples that become unstable, it is likely that the real upper limit is not high enough for this to become a dominant contribution to the population of massive runaways.

\indent Ejected stars tend to escape with velocities on the order of the orbital velocity, and, consequently, triples with more compact inner orbits will produce ejected stars with larger velocities\footnote{Calculating the Pearson correlation coefficient of $\log{v_{\rm esc}}$ as a function of $\log{a_{\rm in, init}}$ gives a value of $-0.74$, showing that there is a clear exponential anticorrelation between the two quantities.
}. Combining this with the observations that massive stars in binaries tend to have compact orbits \citep{sana_binary_2012, moe_mind_2017}, TBZ23 proposes that destabilized triples may play an important role in answering the questions about the origins of OB runaways. In this study we are able to produce high velocity massive runaways, though we can not conclude nor dismiss the idea that triples account for a significant fraction of observed OB runaways due to our initial conditions. Our sample of destabilized triples that eject a star are heavily biased towards wider inner orbits (Fig. \ref{fig:properties_after_escape}), meaning that we are not investigating the outcome of destabilized triples with more compact orbits, which would be the progenitors of the highest velocity stars. There is a combination of factors that cause our population of unstable triples to have wide inner orbits. As seen in \cite{kummer_main_2023}, the distribution of the inner semi-major axis from which the triple systems are sampled peaks at $\sim$$10^3$ R$_\odot$ with an exponentially decaying tail towards higher separations. The outer semi-major axis peaks at $5\times 10^6$ R$_\odot$, with a similar exponential decaying tail to lower values. When generating a triple system from these distributions, there are four general categories that can be initialized: (1) small $a\ssin$ and small $a\ssout$, (2) large $a\ssin$ and small $a\ssout$, (3) large $a\ssin$ and large $a\ssout$, and (3) small $a\ssin$ and large $a\ssout$. Any of the cases with a small $a\ssin$ are more likely to experience inner RLOF before any other interaction occurs. Thus, the  remaining candidates for producing unstable triples from winds alone are the ones with a large inner $a\ssout$. A consequence of the large typical outer SMA means that it is statistically more likely that systems with a wide outer orbit are generated, and as a result, the triples that become unstable from winds must have extremely wide inner orbits if the stellar winds are to expand the orbit enough for instability to occur.

\subsubsection{Runaways from RLOF-induced instability}

Given that the velocity of an ejected star is strongly anti-correlated with the initial inner separation, we investigate what values of $a\ssin$ are required to produce stars with high velocities ($\geq 30$ km/s). Out of the $25 000$ triples simulated by \cite{kummer_main_2023}, $16 833$ experienced RLOF in the inner binary as the first interaction. As mentioned in Section (\ref{sec:instability_rates}), this mechanism might destabilize triples through expansion of $a\ssin$. To get an idea how much the inner separation can expand due to mass transfer, we look at two cases: conservative mass transfer and completely non-conservative mass transfer with isotropic reemission. For these two cases, the change in the orbital configuration is given by \citep{soberman_stability_1997, pols_stellar_1998}:

\begin{equation}
    \frac{a_f}{a_i} = \left( \frac{M_ {\rm d, i}}{M_{\rm d, f}}\frac{M_{\rm a, i}}{M_{\rm a, f}} \right)^2,
\end{equation}

\noindent and,

\begin{equation}
    \frac{a_f}{a_i} = \frac{M_{\rm d, i} + M_{\rm a}}{M_{\rm d, f} + M_{\rm a}} \left( \frac{M_{\rm d, i}}{M_{\rm d, f}} \right)^2  exp{\left(-2\frac{M_{\rm d, i} - M_{\rm d, f}}{M_{\rm a}}\right)},
\end{equation}

respectively. Here, subscript $i$ and $f$ denote the initial and final values respectively, while $d$ and $a$ denote the donor star and the companion. We assume the entire envelope is entirely stripped from the donor, which means that in the case of conservative MT, the final mass of the accretor is equal to its initial mass plus the initial envelope mass of the donor. We then check whether a system would undergo stable or unstable MT using the critical mass ratio $q_{\rm crit}$ from \citet{claeys_theoretical_2014}. Out of the systems that undergo inner MT, $47$ \% of these will be stable. From these systems, $55$ \% of them will increase their separation such that the triple crosses over the instability limit. Finally, $23$ \% of these will destabilize with an inner separation $a\ssin \leq 10^3$ R$_\odot$. We find no significant difference in the results in the case of conservative vs. non-conservative MT. If we assume that out of these systems, $40$ \% eject one of the components, then we find that $1.6$ \% of all massive triples can produce high-velocity runaway stars due to destabilization from inner RLOF, which is 80 times higher than the equivalent rate from stellar winds alone. We emphasize that these results are from back-of-the-envelope calculations, and thus do not encompass the full picture of inner mass transfer in a triple system. For example, we have assumed that the outer orbit does not change during the process. In the case of completely non-conservative mass transfer, the outer orbit is likely to widen due to mass lost from the system. If both the inner and outer orbit widens, the triple might never reach instability, though this depends on the timescales of the expansions, which is beyond the scope of this study. Consequently, these results give an upper limit for the number of system that become unstable due to inner mass transfer. 


\subsubsection{Ejected black holes}

We find that $26$ \% to $42$ \% of the ejected stellar objects are black holes. These tend to move with velocities similar to that of the ejected stars (Fig \ref{fig:escape_velocity_aof_stellar_type}), with slightly higher velocities in the SE model due to the addition of a supernova kick. Single black holes can potentially be observed as a microlensing events \citep{wambsganss_gravitational_2006}. From our Galactic rates, we estimate that the ejection of a black hole from an unstable triple to happen at a rate of $0.3 - 0.5$ per Myr. 

\subsection{Collisions}\label{sec:discussion_collisions}

A significant fraction of the destabilizing massive triples end up with a collision between two main sequence stars. These events are thought to be the progenitors of luminous red novae \citep{kulkarni_unusually_2007, thompson_new_2009, tylenda_v1309_2011}. The rates of massive triples that become unstable and produce a collision correspond to an estimated Galactic rate of $1.1 - 1.3$ events/Myr. The majority of these involve the collision of two MS stars, with the second largest population being that of (post-MS + MS) collisions. Collisions tend to happen in orbits with a small initial inner SMA (Figure \ref{fig:initial_sma_ecc_dist}). In the case of pure gravity, the 80th percentile of collision systems is around $10^5$ R$_\odot$, while this is about one order of magnitude lower when we include SE. In other words, including SE tends to favour smaller inner separations when collisions occur, while the pure N-body model is more uniformly distributed over a larger range of $a\ssin$. One explanation for this is the increase in radii due to the evolution of the stars; with a larger radii, the systems would not need to be kicked up to extreme eccentricites before collision occur, and would therefore produce collisions at earlier times. This is reflected in Figures \ref{fig:duration_nbody} and \ref{fig:duration_SE}, with more collisions occurring after fewer outer orbits in the SE models. 

\subsubsection{Properties of the newly formed binary}

The properties of the newly formed binary following the collision of two components (Fig. \ref{fig:properties_after_collision}) are calculated using two main assumptions: (1) perfect inelastic collision without mass loss, and (2) each collision, regardless of the impact parameter, merges the two components. For the first assumption, hydrodynamical simulations have shown that only a small amount of mass is actually lost during a MS-MS merger \citep{lombardi_stellar_2002, dale_collisions_2006, glebbeek_evolution_2008}. As the majority of the collisions in this study are between two main sequence stars, this assumption is likely to hold for most of the systems. For systems involving a compact object and a star, the star is more likely to be heavily disrupted due to strong tidal forces \citep{regev_hydrodynamic_1987}, potentially forming a disk around the compact object before the two components merge.  With a potentially higher mass loss during the collision, it is more likely that the newly formed object is not bound to the remaining component, meaning that the collision does not form a new binary, but rather two single objects, one of which is a merger product. The second assumption is less realistic, as contact between two stars may lead to three potential outcomes depending on the impact parameter\footnote{The impact parameter is defined such that a value of 0 refers to a head-on collision, and a value 1 of denotes a grazing fly-by, with the edges of the two objects just barely touching.}.

\subsubsection{Collisions between two MS stars}

A collision between two massive MS stars is likely to lead to a merger, with the merger timescale and final properties of the product depending on several factors, including the components masses and the impact parameter. It is possible that an impact parameter close to unity (grazing encounter) might lead to a contact binary-like system. Several properties of contact binaries are still not well understood, including the effects of tides, the transport of energy between the two stars, and the internal structure \citep{shu_various_1980}. As the stars in a contact binary continue to evolve and expand, they might eventually overflow the L2 equipotential, resulting in an outflow of mass and loss of angular momentum, causing a rapid coalescence and merger \citep{pejcha_binary_2016}. Merger products from coalescence might have peculiar properties, such as being rapidly rotating \citep{dale_collisions_2006}, having high helium abundance in the envelope \citep{ivanova_slow_2002}, or being highly magnetized \citep{marchant_evolution_2023}. Merger products from two MS stars can also have the property of being too luminous and blue in comparison to other MS stars in its local environment (\citet{chatterjee_stellar_2013} and references within).

\citet{dale_collisions_2006} performed hydrodynamical simulations of collisions and close encounters of massive main-sequence stars. They find that close encounters are likely to lead to a rapid merger, with timescales of a few tens stellar free-fall times. Encounters with an impact parameter close to unity (grazing encounters) were found to produce a common envelope system, which may result in the envelope being expelled and potentially leading to a merger. The most massive stars were found to be more likely to expel the envelope without merging of the two cores, resulting in a compact binary consisting of stripped stars. Additionally, energy deposited into the envelope of the merger product might cause them to expand their radii by up to two orders of magnitude. Such a swell-up of the merger product can precipitate further interactions in the newly formed binary following the merger, potentially resulting in mass transfer and/or another merger, this time between one of the original triple components and the merger product. Collision products are likely to be rapidly rotating, due to the orbital angular momentum of the progenitors being deposited into the spin of the remnant or in the mass that is lost during the merger. If little mass is lost, the product will be rapidly rotating, making them possible progenitors for hypernovae or gamma-ray bursts \citep{podsiadlowski_effects_2004}. 

\subsubsection{Collisions between an MS and an evolved star}

In addition to MS-MS collision, we also produce a high rate of collisions between an evolved star and a main sequence star. The post-MS stars involved in collisions have a mean mass of 25 M$_\odot$ with a standard deviation of 8.6 M$_\odot$. The MS counterparts have a similar spread, but with a mean mass of 12 M$_\odot$. The outcome of a collision between a MS star and an evolved star depends on factors such as the mass and the exact stellar type of the post-MS components, and the impact parameter. As noted by TBZ22, there are three potential scenarios that may occur: (1) the MS star may pass through the envelope of the evolved star, stripping some of the mass; (2) the two stars merge, and; (3) the two stars form a contact binary through the process of a common envelope scenario, stripping the envelope in the process. In the case that a system such as this does merge, the hydrogen rich core of the main sequence star rejuvenates the evolved component, effectively extending its lifetime. These remaining hard binaries might become potential progenitors for gravitational wave sources, as the small separation can lead to a rapid coalescence once the stars become compact objects.

\subsection{Low-mass vs. high-mass unstable triples}

As this study focuses exclusively on triple systems with a high-mass primary, we present here an overview of the most important distinctions between our results and those from TBZ23 and HPTN22, who mainly looked at low-mass stars.

Our population of high-mass triples on the edge of stability consists of $79$ \% MS stars, with a smaller percentage - $6.6$ \% - post-MS stars (Fig. \ref{fig:stellar_types_distribution}), while TBZ22 find that $30 - 40$ \% of low-mass triples contain a giant (post-MS) primary star at the onset of instability. Low mass stars have weaker winds while on the main sequence compared to high-mass stars, resulting in limited widening of the inner orbit before the primary evolves off the main sequence. High mass stars are therefore more likely to widen the inner orbit to the point of instability before any of the components become post-MS stars, consequently resulting in a higher fraction of MS stars at the onset of instability.

$35 - 40$ \% of our systems end up in a collision between two components (Fig. \ref{fig:all_outcomes}). This is significantly higher than the collision rate for low-mass systems from TBZ22, at $13 - 24$ \% (Fig. 1 in TBZ23), and $14$ \% in HPTN22\footnote{They report a collision percentage of $54$ \%, but highlight that this includes common envelope evolution, in which a star fills its Roche lobe around a companion. The reported rate of $14$ \% comes from \textit{clean collisions}, in which the radii of two MS or CO components overlap.}. In both high-mass and low-mass systems, collisions tend to happen between two MS stars (Fig. \ref{fig:collider_types} and Fig. 1 in TBZ23). The greater collision percentage in the high-mass population is likely a combination of two main factors: the larger radii of high-mass stars, and the initial orbital separation. The inner orbital separation of the high-mass systems are comparable to that of the low mass systems. Having larger radii in the same orbit results in a higher likelihood for collisions to occur due to a greater collisional cross-section.

\subsection{Model caveats}

As with any model, our model has several shortcomings that should be considered when discussing the results. We know what certain systems experience RLOF, either during the hierarchical phase in an eccentric inner binary, or during a close passage during a democratic interaction. In these simulation we only flag these systems for bookkeeping, but it is likely that many of these occurences would result in either mass transfer, mass loss, or both. RLOF in eccentric orbits is still poorly understood, and thus it is difficult to state whether or not the inclusion of accurate mass transfer/loss prescriptions would significantly affect the outcomes of the systems where one of the stars overflows its Roche lobe. It is possible that some of these interactions are similar to partial tidal disruption events, in which a star is partially disrupted following a close encounter with a compact object \citep{wang_partial_2021}.  

Another obvious caveat is how we include stellar evolution. By pre-calculating the evolution of each star individually, we are assuming they evolve in isolation, an assumption that breaks down for the systems that experience RLOF at any point during the simulations, and also for the systems that do not, as some portion of the mass lost from winds would likely be accreted by the companion(s). Mass transfer and wind accretion is likely to change the outcome of some of the triples due to the chaotic nature of systems with strong three-body effects. However, whether the cumulative effect of these changes would significantly affect the overall statistics is unknown. 

Finally, our N-body implementation does not include tidal effects. The inclusion of tides would introduce another energy dissipation term, which could potentially change the outcome of several systems. Given that our sample mostly contains wider binaries, tidal effects would only be significant during close passages in a democratic phase or during the pericenter passage of a highly eccentric triple. It is likely that the inclusion of tidal forces would produce more compact binaries instead of colliders or escapers by shrinking and circularising the inner orbit (\cite{orlov_effect_1996} and references within), effectively moving the systems back to the region of dynamical stability. Though further evolution of the system may again widen the inner orbit and trigger another instability. However, the overall effect on the statistics is unclear and beyond the scope of this study. 


\section{Conclusion}\label{section:conclusion}

In this work we have studied the final outcomes of massive triple systems that become dynamically unstable due to stellar winds. Using a direct N-body approach coupled with a stellar evolution code we simulated a population of triple systems that reside on the edge of stability. We considered two suites of model, one that only takes into account the gravitational interaction between the bodies, and one model which also including stellar evolution. Our results can be summarized as follows:

\begin{itemize}
    \item Even though all the systems simulated in this work are exactly on the limit of dynamical stability, as defined by Equation (\ref{eq:mardling_aarseth}), a substantial fraction takes up to hundreds and even thousands of outer orbital periods before the system disintegrates. This is in line with the observations from TBZ23. This result not only serves as a justification for including stellar evolution in the simulation, but it also emphasises the importance of simulating unstable triples for a longer timescale before categorizing them as remaining hierarchical.
    \item Between $35$ \% and $40$ \% of the simulated triples end in a collision. The collisions happen overwhelmingly between the primary and secondary components, and occurs mostly between two main sequence stars (between $70$ \% and $78$ \% of collisions). These collisions are likely to merge the two components, producing a star with potentially peculiar properties such as being rapidly rotating or highly magnetized. After MS-MS collisions, the most common scenario involves collisions between an evolved star and a MS star. The evolved stars in these collisions are mostly very massive, and the collisions tend to be grazing encounters. These are likely to result in a common envelope-like event, which might either strip the envelope and produce a compact binary, or merge the two components. In the case of a compact binary, these might be progenitors for future gravitational wave sources, while a merger between these two components can rejuvenate the evolved star, producing a blue straggler-like star.
    \item $31$ to $40$ \% of systems eject one of the components, leaving behind a newly formed binary and a single stellar object. The distribution of velocities for these ejected bodies peaks at around $4-6$ km/s, with a tail towards larger velocities. About $4$ \% of the ejected bodies are massive stars with velocities $\geq 30$ km/s, putting them firmly in the regime of massive runaways. We estimate a lower limit on massive runaways produced via the unstable triple channel of about $0.02$ \%, which would correspond to about $0.1$ \% of all the massive runaways. 
    \item The velocity of an ejected body is strongly anticorrelated with the inner separation at time of destabilization. The population of triples studied in this work have mainly wide inner orbits, and it is therefore still possible that triples might contribute more significantly to high velocity runaway stars. Other mechanisms for destabilizing triples might favour more compact orbits, such as mass transfer in the inner binary, which in turn can produce ejected stars with substantially higher velocities. By looking at extreme cases of completely conservative and non-conservative mass transfer, we use analytical formulae to estimate the final properties of the massive triples that undergo mass transfer in the inner binary. We find that this can mechanism can produce up to ten times as many unstable triples compared to winds alone, with a significant fraction having with compact inner orbits and may therefore be potential progenitors for high-velocity runaways.  
    \item When stellar evolution is considered, about $20$ \% of the systems completely ionize, meaning that all three stars become unbound from each other due to multiple internal supernovae. These cases do not produce a remaining binary, but rather three single stellar objects, at least two of which will be stellar remnants.
    \item With our Galactic model, we estimate Galactic rates for collisions, ejections, drifters, and ionized triples. All the rates are on the order of $1$ Myr$^{-1}$, which is predominantly due to the low number of massive stars. 
    \item Collisions and ejections both leave behind a binary system. As collisions tend to happen between the components in the inner binary, the properties of the newly formed binaries follow the distributions of the initial outer orbit, thus produce wide binaries. Similarly, the ejection channel tends to expel the tertiary star, leaving behind the inner binary. Consequently, ejections generally produce tighter binaries with orbital parameters closely resembling that of the inner binary. 
    \item Due to the stronger stellar winds present in massive stars, triples with massive primaries tend to destabilize before evolving off the main sequence.
    \item Compared to low-mass triples, systems with a massive primary are more likely to end up in a collision, up to a factor of $2.7$. The orbital separations of unstable triples differs little between low-mass and high-mass systems, and so the increased radii of massive stars results in a larger collisional cross-section.
\end{itemize}


\begin{acknowledgements}
    ST acknowledges support from the Netherlands Research Council NWO (VIDI 203.061 grant). 
    TB is supported by an appointment to the NASA Postdoctoral Program at the NASA Ames Research Center, administered by Oak Ridge Associated Universities under contract with NASA.
\end{acknowledgements}

%
%


\bibliographystyle{aa}
\bibliography{bibliography_bibtex}
  

\begin{appendix}

\end{appendix}

\end{document}